\documentclass[12pt]{article}
\usepackage{amssymb}
\usepackage{amsmath}
\usepackage{graphicx}

\numberwithin{equation}{section}

\begin{document}
\baselineskip=17.5pt
\begin{titlepage}
\begin{flushright}
{\small UT-HET 040}\\[-1mm]%
{\small KUNS-2271}%
\end{flushright}

\begin{center}
\vspace*{10mm}

{\large\bf Deep Correlation Between Cosmic-Ray Anomaly\\[2mm]%
and Neutrino Masses}%
\vspace*{12mm}

Shigeki Matsumoto$^1$ and Koichi Yoshioka$^2$%
\vspace{4mm}

${}^1${\it Department of Physics, University of Toyama, 
Toyama 930-8555, Japan}\\%
${}^2${\it Department of Physics, Kyoto University, 
Kyoto 606-8502, Japan}
\vspace*{12mm}

\begin{abstract}\noindent%
The anomaly recently reported by the cosmic-ray measurements
suggests that, if explained by the decay of dark matter particle, the
decay source is closely linked up with the leptonic sector of the
standard model. It is observed that, with a simple dimensional
analysis, the lifetime of dark matter for the anomaly is expressed by
the energy scale of neutrino masses. We present two scenarios in which
these two matter at issue (the dark matter width and the tiny neutrino
masses) stem from a single operator involving a gauge-singlet scalar
field.
\end{abstract}

\end{center}
\end{titlepage}

\newpage
\section{Introduction}

The dark matter, which accounts for about 23\% of the energy density
in the present Universe~\cite{WMAP}, has been a great mystery in
astrophysics, cosmology, and particle physics. While various
theoretically-motivated candidates have been discussed for dark matter
particle, its detailed nature is still un-revealed. Among experimental
methods being performed, the flux observations of high-energy cosmic
rays are expected to give important information on the dark matter
which can be an extra source of fluxes through its annihilation or
decay. The recent result from the PAMELA experiment~\cite{PAMELA}
indicates the surplus of positron flux over the expected background in
conventional astrophysics, and some non-standard source of
energetic positrons in our galaxy is needed. The dark matter, which
comes from new physics beyond the Standard Model (SM), has been
examined as one of possible explanations of this anomalous behavior,
though it was pointed out that the data may be fitted by the effect of
nearby pulsars~\cite{pulsar}. Subsequent comprehensive analysis
suggests that the cosmic-ray anomaly can be well described if there
exists a dark matter which decays mainly into charged leptons and has
the width $\Gamma_\text{DM}\simeq10^{-26}/\text{sec}$~\cite{analysis}.

Another important problem in the SM is the origin of neutrino mass,
much lighter than the other SM fermions. For Majorana neutrinos, the
seesaw mechanism~\cite{seesaw} naturally works to have tiny masses by
the integration of high-energy degrees of freedom, which leads to a
higher-dimensional operator suppressed by a heavy scale. The neutrino
experimental data implies that the heavy mass scale is around the
grand unification scale to have light Majorana masses of order
eV\@. Alternatively, for Dirac neutrinos, there exist gauge-singlet
fermions in low-energy theory and couple to the SM neutrinos through
the Dirac mass operator. In this case, the mass scale should be
suppressed by some mechanism, e.g.\ in a similar way to the mass
hierarchy of SM fermions for which various approaches have been
proposed in the literature.

In this letter, we point out a deep correlation between these two
problems: the origins of dark matter and neutrino mass. In particular,
we present two scenarios where the long lifetime of dark matter is
directly determined by the smallness of neutrino masses. The dark
matter decay and neutrino mass come from a single operator in
low-energy effective theory, and therefore they are naturally 
related. The close connection to the neutrino sector is also plausible
for the experimental result that the excess is observed for the
positron fraction and not for other components in cosmic rays. We
focus, in this letter, on the implication for particle physics and do
not discuss detailed predictions of the cosmic-ray spectrum, as it is
influenced by various astrophysics factors such as the propagation of
cosmic rays in our galaxy~\cite{propagation}.

\bigskip

\section{Scaling relations for $\Gamma_\text{\rm DM}$}

The unstable dark matter is a reasonable explanation for the
cosmic-ray anomaly~\cite{decayingDM_models}: the lifetime is much
longer than the age of the Universe and the main decay channel
contains positrons and electrons, not colored particles. In addition,
a precise measurement of the  total $e^++e^-$ flux has been performed
by the Fermi Gamma-ray Space Telescope~\cite{FermiLAT_ee}. The
observed positron flux may be significantly contaminated by that from
decaying dark matter with a few TeV mass and the lifetime 
of ${\cal O}(10^{26})$~sec~\cite{analysis}. It is interesting to
notice that these two physical numbers implies a scaling relation for
the decay width of TeV-scale dark matter:
\begin{eqnarray}
  \Gamma_\text{DM} \;\simeq\; 10^{-26}/\text{sec} \,\;\simeq\;
  \bigg(\frac{1\,\text{TeV}}{\Lambda}\bigg)^4\,\text{TeV},
  \label{relation1}
\end{eqnarray}
with $\Lambda\,\simeq\,10^{16}$~GeV that indicates the existence of
unification-scale dynamics behind the dark matter data. For instance,
the above relation follows if there exists a fermionic dark matter
particle with TeV mass and it decays to the SM sector through a
4-Fermi operator suppressed by $1/\Lambda^2$. Motivated by this
observation, the phenomenological analysis and various construction of
grand unified models ($\Lambda=M_\text{GUT}$) have been
performed~\cite{decayingDM_GUT}.

If the neutrino physics has some relevance to the dark matter, another
interesting viewpoint is found. Namely, noticing that a typical mass
scale of neutrinos indicated by oscillation experiments 
is $m_\nu\simeq10^{-1}$~eV and 
numerically $(m_\nu/\text{TeV})\simeq(\text{TeV}/M_\text{GUT})$, the
dark matter decay width in \eqref{relation1} has a different expression
\begin{eqnarray}
  \Gamma_\text{DM} \;\simeq\;
  \bigg(\frac{m_\nu}{\Lambda}\bigg)^2\,\text{TeV},
  \label{relation_neu}
\end{eqnarray}
where $\Lambda$ is around the unification scale. The new scaling
relation~\eqref{relation_neu} suggests that, in low-energy theory
below $\Lambda$, there is an effective operator suppressed 
by $1/\Lambda$ which causes the dark matter decay and is closely
related to the neutrino mass in light of the cosmic-ray 
anomaly.\footnote{There are other attempts on joining together the
neutrino masses and the cosmic-ray anomaly from decaying dark
matter~\cite{neutrinoDM}.}

Let us clarify the property of dark matter from which the
relation~\eqref{relation_neu} is deduced. In the SM extension with a
dark matter field $X$, there are four types of dark matter depending
on the statistics and electroweak charge. We assume that the TeV-mass
dark matter couples to the SM lepton doublets $L$ through an effective
operator relevant for neutrino masses. An electroweak singlet fermion
such as right-handed neutrinos $\nu_R$ is an immediate candidate for
this property. An non-singlet (and charge neutral) fermion can also
couple to $L$ with a (composite) scalar in a higher-dimensional
representation of electroweak symmetry. In both cases, one easily
finds with a dimensional counting that the decay width of $X$ is
roughly estimated as $\Gamma_X\sim m_\nu$, instead of $m_\nu^2$ given
in \eqref{relation_neu}, if the gauge invariant 
operator $\bar{L}X{\cal O}$ is relevant for neutrino masses 
where ${\cal O}$ contains the SM Higgs and other scalar fields. Such a
large width is not suitable for the cosmic-ray anomaly.

Another candidate of dark matter $X$ is a TeV-mass scalar field. With
an appropriate hypercharge, it can contain a neutral component as the
dark matter~\cite{multiplet}. Both for Dirac and Majorana neutrinos,
their masses are transformed non-trivially under the SM gauge symmetry
and come from the operators $\bar{L}\nu_R{\cal O}'$ 
and $\bar{L^c}L{\cal O}''$ supplemented by vacuum expectation values
(VEV) of $X$ and/or the SM Higgs, where ${\cal O}'$, ${\cal O}''$ are
(composite) scalar operators. Expanded around the vacuum, the
effective scale $\Lambda$ in this case is given by these VEVs which
are at most the electroweak scale. That clearly leads to a rapid decay
of dark matter. On the other hand, when $X$ does not develop its VEV,
the neutrino mass operators given above should be generated by
integrating out the heavy $X$. In this case, the decay 
width $\Gamma_X$ is found to be larger than $m_\nu$ and not suited for
the cosmic-ray anomaly.

In the end, the electroweak singlet scalar is found to be a reasonable
candidate for the decaying dark matter with its width being given by 
Eq.~\eqref{relation_neu}. The singlet scalar $\phi$ has two important
properties for the scaling relation to work: (i) it decays into the SM
sector through a higher-dimensional operator (and other decay vertices 
are suppressed) and (ii) it develops a non-vanishing VEV which
induces neutrino mass from that operator. The dark matter filled in
our Universe is the quantum fluctuation about the VEV:
\begin{eqnarray}
  \phi \;=\; \langle\phi\rangle+\phi_\text{DM}.
  \label{DMvev}
\end{eqnarray}
The cosmic-ray anomaly gives us a signal that $\langle\phi\rangle$ is
around the unification scale and $\phi_\text{DM}$ has a TeV-scale mass
in the vacuum. That is not unnatural, e.g.\ in supersymmetric theory
which would help the dark matter scalar to realize a stabilized flat
potential for a large VEV and to receive soft supersymmetry breaking
for a small mass. Another way to incorporate a tiny mass/VEV ratio is
to consider the dark matter as a Nambu-Goldstone field, as will be
discussed later. In any case, the dark matter couples to the lepton
sector via an effective operator for neutrino masses and the excess of
positron fraction is naturally explained.

\bigskip

\section{Decay from neutrino mass}

There are two cases (two effective operators) for neutrino masses
where the relation~\eqref{relation_neu}, i.e.\ the dark matter
property described above, is encoded. The types of effective operators 
depend on whether neutrinos are Dirac or Majorana particles. While the
present status of neutrino experiments does not discriminate between
these operators, different theoretical frameworks beyond the SM are
constructed according to it. 

We first consider the Dirac neutrinos, namely introduce light
right-handed neutrinos $\nu_R$ ($\nu_R=P_R\nu_R$) in addition to the
dark matter scalar $\phi$. This model has the lepton number symmetry
even after the condensations of dark matter and Higgs fields. The
gauge-invariant operator for neutrino masses is given by 
\begin{eqnarray}
  \text{\underline{Dirac}} : \qquad
  {\cal L}_D \;=\; -\frac{y}{M}\,\phi\,\tilde H^\dagger
  \overline{\nu_R} L
  \,+\text{h.c.}, \qquad
  \label{L_Dirac}
\end{eqnarray}
where $H$ is the SM Higgs field ($\tilde H=\epsilon H^*$) and $L$ is
a lepton doublet ($L=P_LL$). The $\phi$ dependence could be fixed by
symmetry argument and/or high-energy dynamics at $M$, but its detail
is found to be almost irrelevant to the dark matter physics (see
below). Inserting the expectation values 
of $\phi$ and $H$ $\,[\,$Eq.~\eqref{DMvev} 
and $\langle H\rangle\!=\!(0,v/\sqrt{2}\,)^\text{t}\,$], we obtain
the Dirac mass $m_\nu$ for neutrinos $\nu=\nu_L+\nu_R\,$; 
\begin{eqnarray}
  m_\nu \;=\; \frac{yv\langle\phi\rangle}{\sqrt{2}M}.
\end{eqnarray}
Expanded around the vacuum, the Lagrangian~\eqref{L_Dirac} is
rewritten as
\begin{eqnarray}
  {\cal L}_D \;=\; -m_\nu\Big(1+\frac{h}{v}\Big)\bar\nu\nu
  -\frac{m_\nu}{\langle\phi\rangle}\Big(1+\frac{h}{v}\Big)
  (\phi_R\,\bar\nu\nu -\phi_I\,\bar\nu i\gamma_5\nu),
  \label{L_Dirac2}
\end{eqnarray}
where $h$ is the Higgs boson and $\phi_\text{DM}=\phi_R+i\phi_I$. We
have taken $\langle\phi\rangle$ and $m_\nu$ to be real with suitable
phase rotations. It is interesting to find from this Lagrangian that
the dark matter decay is governed by only two 
quantities, $\langle\phi\rangle$ and the dark matter 
mass $m_\text{DM}$, and is independent of the model 
parameters $y$ and $M$ which are just utilized to have a proper scale
of neutrino masses. This fact holds, namely the
Lagrangian~\eqref{L_Dirac2} is almost invariant, even if the dark
matter dependence in the operator~\eqref{L_Dirac} is replaced with an
arbitrary function $Y(\phi)$. All decay amplitudes are multiplied by a
single common factor $\partial\ln Y(\langle\phi\rangle)/
\partial\ln\langle\phi\rangle$. 

The dominant decay of dark matter occurs through the second term
in~\eqref{L_Dirac2} and is given by the (lepton number 
conserving) 2 and 3-body decays at tree 
level: $\phi_\text{DM}\to\nu\bar\nu$ and $\nu\bar e W,\,
\nu\bar\nu Z,\,\nu\bar\nu h$. The decay vertices are fixed by the
ratio $m_\nu/\langle\phi\rangle$ and the expectation 
value $\langle\phi\rangle$ plays the role of $\Lambda$ in the scaling
relation discussed in the previous section. We find the partial decay
widths
\begin{eqnarray}
  \qquad
  \Gamma_{\nu\bar\nu} \;=\;
  \frac{m_\text{DM}^{}}{8\pi}\frac{m_\nu^2}{\langle\phi\rangle^2}\,, 
  \qquad\;
  \Gamma_{\nu\bar eW}\simeq
  \Gamma_{\nu\bar\nu Z}\simeq\Gamma_{\nu\bar\nu h} \simeq\, 
  \frac{m_\text{DM}^3}{768\pi^3v^2}
  \frac{m_\nu^2}{\langle\phi\rangle^2}\,.
  \label{width}
\end{eqnarray}
For the latter formula, we have dropped the sub-leading terms
suppressed by $v^2/m_\text{DM}^2$. In the SM language, the above
chirality-violating decays are caused by the neutrino Yukawa and
electroweak dipole operators accompanied by the dark matter
scalar. The 3-body decays are comparable to the 2-body one, since the
phase space suppression is supplemented with the enhancement 
factor $m_\text{DM}^2/v^2$. The decay widths become 
\begin{eqnarray}
  \Gamma_{\nu\bar\nu} \;\simeq\; 1.8\times10^{-26}/\text{sec}, \qquad
  \Gamma_\text{3-body} \;\simeq\; 0.3\times10^{-26}/\text{sec},
\end{eqnarray}
for typical values $m_\nu=10^{-1}\>$eV, $m_\text{DM}=3\>$TeV,
and $\langle\phi\rangle=10^{16}\>$GeV\@. The positrons are emitted by
one of the 3-body decays, $\phi_\text{DM}\to\nu\bar eW^-$, and the
cosmic-ray anomaly observed by the PAMELA experiment indicates the
dark matter with a few TeV mass and an expectation 
value $\langle\phi\rangle$ of the unification scale. That is exactly
what we infer from the scaling relation~\eqref{relation_neu}.

\bigskip

Another case is the Majorana neutrino. The seesaw mechanism 
leads to tiny neutrino masses by integrating out heavy right-handed
fermions. The Lagrangian relevant for neutrino masses is given by
\begin{eqnarray}
  \text{\underline{Majorana}} : \qquad
  {\cal L}_M \;=\; -y\tilde H^\dagger\overline{\nu_R} L
  -\frac{f}{2}\,\phi\,\overline{\nu_R^{\,c}}\nu_R^{} 
  \,+\text{h.c.},  \qquad
  \label{L_Majorana}
\end{eqnarray}
where $y$ is the neutrino Yukawa coupling and $\nu_R^{\,c}$ denotes
the charge-conjugate spinor. Similar to the Dirac neutrino case, the
couplings $y$ and $f$ are not directly relevant to the dark matter
physics. This model has the conserved lepton number if a suitable
charge is assigned to the dark matter scalar $\phi$ and it is broken
in the vacuum.\footnote{The Nambu-Goldstone boson from broken lepton
number symmetry has been discussed as a dark matter candidate in
different contexts~\cite{MajoronDM}.} The expectation 
value $\langle\phi\rangle$ gives the Majorana mass of right-handed
neutrinos: $m_R=f\langle\phi\rangle$. The integration of heavy 
modes ($m_R\gg m_\text{DM}$) induces an effective operator between the
dark matter and SM fields\footnote{If right-handed neutrinos are
lighter than the dark matter, the decay width is proportional 
to $f^2$ or $fm_\nu/\langle\phi\rangle$, depending on $m_R$. For both
cases, the cosmic-ray anomaly implies a too small $f$, in other words,
a trans-Planckian value of $\langle\phi\rangle$ is needed 
unless $m_R<(M_\text{Pl}/M_\text{GUT})m_\nu$.}
\begin{eqnarray}
  {\cal L}_M \;=\; 
  \frac{y^Ty}{f^*\phi^*}\,\overline{L^c}\tilde H^*\tilde H^\dagger L
  +\text{h.c.}\,.
  \label{L_Majorana2}
\end{eqnarray}
Inserting the expectation values of $\phi$ and $H$, we obtain the
seesaw-induced Majorana mass $m_\nu$ for light 
neutrinos $\nu=\nu_L+\nu_L^{\,c}\,$;
\begin{eqnarray}
  m_\nu \;=\; -\frac{y^Tyv^2}{m_R^*}.
\end{eqnarray}
Expanded around the vacuum, the Lagrangian~\eqref{L_Majorana2} is 
rewritten as
\begin{eqnarray}
  \qquad
  {\cal L}_M \;=\; -\frac{1}{2}m_\nu\Big(1+\frac{h}{v}\Big)^2\bar\nu\nu
  +\frac{1}{2}\frac{m_\nu}{\langle\phi\rangle}\Big(1+\frac{h}{v}\Big)^2
  (\phi_R\,\bar\nu\nu +\phi_I\,\bar\nu i\gamma_5\nu)
  \,+{\cal O}(\phi_\text{DM}^2),
  \label{L_Majorana3}
\end{eqnarray}
where $h$ is the Higgs boson and $\phi_\text{DM}=\phi_R+i\phi_I$. We
have assumed that $\langle\phi\rangle$ and $m_\nu$ are made real by
phase rotations. Notice that the Lagrangian is very similar to that in
the Dirac neutrino case, except that $\nu$ is a Majorana particle. The
dark matter decay via the operator~\eqref{L_Majorana2} is governed 
by $\langle\phi\rangle$ and $m_\text{DM}$, and is independent of the
Yukawa couplings. This fact holds even if the dark matter dependence
in~\eqref{L_Majorana} is replaced with an arbitrary 
function $F(\phi)$. All decay amplitudes with the seesaw operator are
multiplied by a single common factor $\partial\ln F(\langle\phi\rangle)/
\partial\ln\langle\phi\rangle$.

The dark matter decay through the operator~\eqref{L_Majorana2} occurs
in the exactly same fashion as the Dirac neutrino case with the
replacement $\nu_R\to\epsilon\nu_L^*$. The second terms
in~\eqref{L_Majorana3} therefore lead to the (lepton number 
violating) 2 and 3-body decays at tree 
level: $\phi_\text{DM}\to\nu\bar\nu$ and 
$\nu\bar e W,\,\nu\bar\nu Z,\,\nu\bar\nu h$, and the partial decay
widths are respectively given by Eq.~\eqref{width}. The decay vertices
are fixed by the ratio $m_\nu/\langle\phi\rangle$ and the expectation
value $\langle\phi\rangle$ is around the unification scale so that the
scaling relation is concluded irrespectively of other coupling
constants. These chirality-violating decays are brought about by the
Yukawa and electroweak dipole operators. On the other hand, there are
several chirality-conserving operators generated at 1-loop 
level.\footnote{For the Dirac neutrino case, such operators are not
induced to the extent that the tree-level result is affected, because
of tiny chirality-flip mass parameters.} 
They include $\phi_\text{DM}|H|^2$, $\phi_\text{DM}|D_\mu H|^2$, and
$\phi_\text{DM}\bar L\gamma^\mu D_\mu L$, where $D_\mu$ means the
electroweak covariant derivative. The former two types of operators
induce the decays into two electroweak 
bosons, $\phi_\text{DM}\to W^+W^-,ZZ,\,hh$, and tend to be the main
decay modes. These decays would cause an undesirable anti-proton
excess in cosmic ray as well as positrons but are all forbidden 
by identifying the imaginary part $\phi_I$ with the dark matter. The
last leptonic operator gives rise to the decay into a pair of charged
leptons or neutrinos whose amplitude, due to the chirality
conservation, receives a helicity suppression to an unobservable
level. The last operator also induces chirality-preserving 3-body
decays, e.g.\ $\phi_\text{DM}\to\nu\bar eW$. While they are suppressed
by the 1-loop factor of electroweak gauge coupling, the necessary
chirality flip in the loop gives an enhancement 
factor $m_R/\text{TeV}$ which makes the loop contribution important
for heavier right-handed neutrinos. The partial width of loop-level
decay is roughly given 
by $\Gamma_\text{loop}\sim (m_\nu/\langle\phi\rangle)^2 
m_\text{DM}^5m_R^2/(8\pi^2v^2)^3$ and would be dominant 
for $m_R\gtrsim{\cal O}(10)$~TeV\@.

\bigskip

We finally mention a hybrid of the two models: the Lagrangian for
neutrino masses and dark matter scalar is given by
\begin{eqnarray}
  \text{\underline{Hybrid}} : \qquad
  {\cal L}_H \;=\; -\frac{y}{M}\,\phi\,\tilde H^\dagger\nu_R^\dagger L
  +\frac{m_R}{2}\,\nu_R^{T}\,\epsilon\nu_R^{}  \,+\text{h.c.}.
\end{eqnarray}
For a negligible order of $m_R$, this model reduces to the Dirac
neutrino case. In another situation with $m_R\gtrsim{\cal O}(1)$~eV,
the seesaw operation should be viable and the model is similar to the
Majorana neutrino case. The only difference is that, for 
lighter $\nu_R$ than the dark matter, the decay width is suppressed 
by $m_\nu/m_R$ compared with the Majorana case. This fact could 
help the model by decreasing the parameter region of a huge value 
of $\langle\phi\rangle$.

\bigskip

Several comments on phenomenology and high-energy dynamics are made in
order:\\[1mm]
(i) \underline{Leptonic decay} :
The cosmic-ray observation suggests that the dark matter interaction is
leptonic and the scaling relation implies that the dark matter decay
is neutrino-involving. That is clearly seen in the Majorana neutrino
case: the SM gauge invariance ensures that the dark matter scalar can
only couple to the matter sector via right-handed neutrinos at the
renormalizable level. The lepton number symmetry also provides this
picture with an attractive support. As for the Dirac neutrino case, an
extra (discrete) symmetry might be needed not to have an anomalous
excess of anti-proton flux. For example, the dark matter and
right-handed neutrinos are charged under such a symmetry so that the
combination $\phi^*\nu_R$ becomes singlet. An interesting possibility
is the higher-dimensional theory in which only right-handed neutrinos
propagate in the bulk~\cite{bulk_nuR} and the dark matter is supplied
by a moduli field associated with the extra-dimensional space such as
the radion. The translational invariance ensures that it couples to
the SM sector only through the neutrino Yukawa term (the leading
bulk-boundary interaction) and hence decays into the leptonic
modes. Furthermore the neutrino mass and the decay rate of dark matter
could be made tiny by the volume suppression factor from the extra space.
\\[2mm]
(ii) \underline{Dark matter symmetry} :
The effective mass operator~\eqref{L_Dirac} can be described in a
similar way to the other SM fermions masses. For example, it is
extended to a polynomial form $Y(\phi)=y(\phi/M)^n$ (and hence the
decay width is multiplied by $n^2$). The exponent $n$ is determined by
high-energy dynamics above $M$ such as the Froggatt-Nielsen 
mechanism~\cite{FN} with $U(1)$ symmetry under which the dark matter
scalar is charged. Such a symmetry is also responsible for avoiding a
rapid decay of this Froggatt-Nielsen dark matter via a coupling to the
SM Higgs of the form $\phi|H|^2$. Furthermore, if the dark matter is
identified with the imaginary component $\phi_I$, its Nambu-Goldstone
nature from the broken $U(1)$ symmetry makes it natural to have a much
smaller mass than the (unification-scale) expectation value. Such
property of dark matter is also true for the Majorana neutrino case:
the lepton number symmetry plays the same role as the Froggatt-Nielsen
symmetry. One may wonder whether, if such symmetry were anomalous,
that induces a loop-level decay of dark 
matter $\phi\to W^+W^-$ etc.\ through a dimension-five anomaly
operator $\epsilon^{\mu\nu\rho\sigma}\phi 
D_\mu W_\nu^+D_\rho W_\sigma^-$ which would lead to an excess of the
anti-proton flux as well as the positron. However this is not the case
for the present two scenarios: the anomaly operator coefficient is 
suppressed by small left-right mixing and becomes negligible. For the
Dirac neutrino case, the partial width of anomalous decay is
proportional to $m_\nu^4/\langle\phi\rangle^2$ and for the seesaw
case, the amplitude is found to be proportional 
to $m_\nu/m_R^2$, i.e.\ the decay occurs through a dimension-seven
operator $\epsilon^{\mu\nu\rho\sigma}\phi(D_\mu D_\nu H)^\dagger
D_\rho D_\sigma H$ in low-energy effective theory.
\\[2mm]
(iii) \underline{Relic abundance} :
A gauge-singlet scalar has a renormalizable interaction to the SM
sector via $|\phi|^2|H|^2$. (As mentioned above, the $\phi|H|^2$ term
is absent if $\phi$ has a nonzero charge of some symmetry.) \ In
appropriate regime of parameter space, it provides a natural
explanation of the relic abundance of dark matter in the present
Universe~\cite{abundance}. Since $|\phi|^2=
(\phi_R+\langle\phi\rangle)^2+\phi_I^2$, the imaginary component is
interpreted as the dark matter and the real part has already decayed
in the early Universe.

\bigskip

\section{Summary}

To summarize, we have pointed out a deep connection between the
neutrino and dark matter physics in light of the recently observed
cosmic-ray anomaly and presented two scenarios, as it were, the
Froggatt-Nielsen dark matter and the seesaw dark matter, where a
single effective operator involving the gauge-singlet scalar produces
correlated sizes of neutrino masses and the dark matter
lifetime. While the form of effective operator and typical scales of
coupling constants would be dictated by high-energy dynamics, its
details are almost irrelevant to the dark matter decay.

The scenarios described in this paper have an important prediction on
future neutrino telescopes~\cite{neu_flux}. When the dark matter mass
is $O(1)$~TeV, the neutrino pair production is the main decay mode and
its branching ratio is about ten times larger than that for charged
leptons. This will be an smoking-gun signature to confirm that the
cosmic-ray anomaly comes from the decay of dark matter, since the
signal is observed as a neutrino flux with a monochromatic
spectrum. It is also important to evaluate the anti-proton and
gamma-ray fluxes caused by the dark matter decay from which the weak
gauge bosons are produced in addition to the charged leptons. A recent
analysis~\cite{eW} of similar decay modes such 
as $\text{DM}\to eW$ and $\text{DM}\to\mu W\to e\nu\bar\nu W$ shows
that, depending on the diffusion parameters, the scenario would be
constrained by the measurements of anti-proton flux at the
PAMELA~\cite{PEMALA_anti-p} and the diffused gamma ray at the
Fermi-LAT~\cite{FermiLAT_diffused_gamma}. These astrophysical
examination are important for corroborating the scenarios and the
scaling relation.

\bigskip\bigskip

\subsection*{Acknowledgments}
\noindent
This work is supported by the scientific grants from the ministry of
education, science, sports, and culture of Japan (No.~20740135,
21740174, 22244021), and also by the grant-in-aid for the global COE
program "The next generation of physics, spun from universality and
emergence" and the grant-in-aid for the scientific research on
priority area (\#441) "Progress in elementary particle physics of the
21st century through discoveries of Higgs boson and supersymmetry"
(No.~16081209).

\newpage

\end{document}